\documentclass[12pt]{article}
\usepackage{mathrsfs,amsmath,amsfonts,amsbsy}
\begin{document}


\title{\vskip1cm
Finite-size effects on a lattice calculation
\vskip.5cm}
\author{Rafael G. Campos and Eduardo S. Tututi\\ 
Facultad de Ciencias F\'{\i}sico--Matem\'aticas, \\
Universidad Michoacana \\
58060 Morelia, Michoac\'an, M\'exico\\[.5cm]
{\tt\small rcampos@umich.mx,   tututi@umich.mx}\\ 
}
\date{}
\maketitle
{
}\\[1cm]

\begin{center} Abstract \end{center}
We study in this paper the finite-size effects of a non-periodic lattice on a lattice calculation. To this end we use a finite lattice equipped with a central difference derivative with homogeneous boundary conditions to calculate the bosonic mass associated to the Schwinger model. We found that the homogeneous boundary conditions produce absence of fermion doubling and chiral invariance, but we also found that in the continuum limit this lattice model does not yield the correct value of the boson mass as other models do. We discuss the reasons for this and, as a result, the matrix which cause the fermion doubling problem is identified.
\vfill
\newpage
\section{Introduction}\label{intro}
In addition to large computing resources, numerical lattice calculations require of a detailed analysis of finite lattice models to support the numerical calculations in order to have control over the finite-size effects. The results obtained by numerical calculations using finite lattice models of quantum-field problems may not be completely correct if these effects are not taken into account \cite{Fuk92}.\\
In a previous paper \cite{Cam07} we have considered the Schwinger model on a finite and non-periodic lattice consisting of zeros of the Hermite polynomials that incorporates a well-behaved lattice derivative and a discrete Fourier transform. We found that if an infinite number of sites are taken into account, the mass of the boson mode calculated analytically by using this lattice becomes the correct one, i.e, $e/\sqrt{\pi}$, but it is about 40\% less than this value if a finite lattice is used in the calculations.\\ 
Good agreement of some finite-lattice calculations with continuum results has been reported elsewhere \cite{Aro99, Jan92, Ell87, Pan08}. Besides the use of different techniques in these papers, it is also worth to note the use of periodic lattices or periodic boundary conditions, which are conditions usually  assumed in lattice theory.\\
Our aim in this letter is to study the finite-size effects of a non-periodic model on a lattice calculation. To this end we calculate the bosonic mass associated to the Schwinger model by using a naive lattice model consisting in a set of $N$ evenly spaced points and a standard central difference derivative with homogeneous boundary conditions. We choose homogeneous boundary conditions because the quantum fields are expected to vanish at $\pm\infty$, whereas periodic boundary conditions are more natural in bounded manifolds. 

The use of homogeneous boundary conditions introduces important changes with respect to the lattice model with periodic boundary conditions. Chirality is maintained with no fermion doubling. The Nielsen-Ninomiya theorem \cite{Nie81} does not apply since the lattice discretization used here is not translationally invariant. However, the continuum value of the boson mass for the one-flavor Schwinger model can not be obtained.
\\
\section{The standard derivative}\label{deriva}
Let the lattice sites $x_n$ be defined in $(-L,L)$ by $x_n=[-1+ 2 n/(N+1)]L$, $n=1,\ldots,N$. The lattice spacing is $a=2 L/(N+1)$ and the boundary points are obviously, $x_0=-L$ and $x_{N+1}=L$. The lattice length should be an increasing function of $N$. It is convenient to choose the dependence
\begin{equation}\label{ndepleng}
L=\pi\sqrt{\frac{N}{2}},
\end{equation}
as we show below. Since a continuum problem is discretized only at the lattice sites, the homogeneous boundary conditions allow us to write the elements of the differentiation matrix $D$ as
\begin{equation}\label{diffinsim}
D_{jk}=\frac{1}{2a}(\delta_{j+1,k}-\delta_{j,k+1}),\quad j,k=1,2,\ldots,N.
\end{equation}
Eventually, it will be necessary to have a diagonal form of this derivative. The usual Fourier  transformation can not be used here, since we are not using periodic boundary conditions. Thus, the first thing to do is to obtain a diagonalizing matrix $F$ for the derivative (\ref{diffinsim}). Because the Fourier transform diagonalizes the derivative, this matrix will be the discrete Fourier transform for this problem. It can be obtained by solving the eigenvalue problem $DV_k=ip_k V_k$, $k=1,2\ldots,N$. Let $v_n(p_k)$ denote the $n$th component of the vector $V_k$. Making the transformation $u_n(p_k)=i^nv_n(p_k)$, the eigenvalue problem for $D$ converts into the finite recurrence equation
\[
u_{n+1}(p_k)+u_{n-1}(p_k)=p_ku_n(p_k).
\]
This is the recurrence equation for the Tchebichef polynomials. Therefore $u_n(p_k)$ is proportional to one of these polynomials. The normalization constant can be found by using the Christoffel-Darboux formula \cite{Cam08}. Thus, the unitary matrix $F$ that diagonalizes $D$ and solves 
\begin{equation}\label{eigprobun}
DV_k=ip_k V_k
\end{equation}
is given by
\begin{equation}\label{vnk}
(V_k)_q=F_{qk}=i^{q-1}\sqrt{\frac{2(1-p_k^2)}{N+1}}\frac{U_{q-1}(p_k)}{\vert U_{N-1}(p_k)\vert},\quad q=1,2,\ldots,N,
\end{equation}
where $U_n(\xi)$ is the $n$th Tchebichef polynomial of the second kind. The eigenvalues of $-iD$ are the momenta $p_k=\xi_k/2a$, where $\xi_k$ is the $k$th zero of $U_N(\xi)$ and therefore, they are not repeated. Thus, if we denote by $P$ the diagonal matrix whose nonzero elements are $p_1,p_2,\ldots,p_N$,  we have that
\begin{equation}\label{fddfp}
F^\dagger (-iD) F=P.
\end{equation}
It is important to notice that the lattice model used here is not translationally invariant and that the lattice momentum $P$ does not  lay on a torus. Thus, the Nielsen-Ninomiya theorem does not apply and this model does not exhibit fermion doubling. In fact, any pair of matrices $(F,D)$ satisfying a one-to-one relation like (\ref{fddfp}) gives a lattice model free from doublers and chirally invariant (if the lattice is not translationally invariant), which is also the case of the lattice studied in \cite{Cam01} (see also Sec 4. of \cite{Cam02}). Below, in Sec. \ref{secdis} we show that the differentiation matrix which is related to the fermion doubling problem is a nearest-neighbor-like derivative $\Delta$, which is actually a generator of forward-backward rotations and it is indeed diagonalized by the discrete Fourier transform \cite{Cam01b}.\\
Since $F$ is unitary, the set of eigenvectors given in (\ref{vnk}) for $k=1,2,\ldots,N$, can be used to expand any other vector in its $F$-components. For $N$ great enough and $k$ fixed, the form of (\ref{vnk}) is
\begin{equation}\label{vnkas}
F_{qk}=(-i)^{q-1}\sqrt{\frac{2}{N}}\sin\frac{qk\pi}{N},
\end{equation}
where $q,k=1,2\ldots,N$. Note that $q$ is the index associated to the position $x_q$ and $k$ is the momentum index. A more useful form is
\[
F_{qk}=\frac{1}{\sqrt{2N}}\left(\exp\left[ix_q\frac{\pi}{a}\left(\frac{k}{N+1}-\frac{1}{2}\right)\right]-\exp\left[-ix_q\frac{\pi}{a}\left(\frac{k}{N+1}+\frac{1}{2}\right)\right]\right),
\]
where $x_q=qa$. In order to take into account the sign of $p_k$, the index $k$ should be changed to $\tilde{k}+(N+1)/2$, where $\tilde{k}=-(N-1)/2,\ldots,(N-1)/2$. This change yields
\begin{equation}\label{vnkasdos}
F_{qk}=\frac{1}{\sqrt{2N}}\left(\exp\left[ix_q\frac{\pi}{a}\frac{\tilde{k}}{N+1}\right]-\exp\left[-ix_q\frac{\pi}{a}\left(\frac{\tilde{k}}{N+1}+1\right)\right]\right).
\end{equation}
For a very large number of points $x_q$ located symmetrically about the origin,
\begin{equation}\label{prodfs}
\frac{1}{2\pi}\sum_q e^{\pm i x_q P}\Delta x_q\to\delta(P).
\end{equation}
In our case, $\Delta x_q=a=2L/(N+1)$. Taking into account (\ref{ndepleng}), we obtain that $1/\sqrt{2N}\simeq\Delta x_q/2\pi$ and therefore,
\[
\sum_q F_{qk}\to \left[\delta\left(\frac{\kappa \pi}{a}\right)-\delta\left(\frac{(\kappa+1)\pi}{a}\right)\right],
\]
or in short, 
\begin{equation}\label{fasfg}
\sum_q F_{qk}\to \left(\delta_{\kappa,0}-\delta_{\kappa,-1}\right),
\end{equation}
where we have defined 
\[
\kappa=\tilde{k}/(N+1), \quad \tilde{k}=-(N-1)/2,\ldots,(N-1)/2.
\]
 This formula will be used below.
\section{The lattice model}\label{latmod} 
We take a lattice Schwinger model based on the Heisenberg equations and canonical quantization, such that the space variable is discretized and time is considered to remain continuous. Once the differentiation matrix $D$ has been chosen and the diagonalizing matrix $F$ has been found, the discretization of the Schwinger model can be done along the same lines as in \cite{Cam07}. As a consequence, many expressions involving $D$ or $F$ obtained in Sec. 2 of \cite{Cam07} remain the same in this case. We are considering a lattice consisting of a very large but {\it finite} number $N$ of points. This means that the sums yielded by the discretization of the integrals are finite sums that can be written in matrix form. Thus, if the number of nodes $N$ is an even integer, the differentiation matrix $D$ is nonsingular and the lattice hamiltonian $H_L$ can be written in the form 
\begin{equation}\label{hamildisfree}
H_L=\frac{1}{2}\rho^t D^{-2}\rho+i \Psi^\dagger(\sigma_z \otimes D) \Psi,
\end{equation}
where $\rho$ and $\Psi$ are related by Gauss' law 
\[
\rho=e\, {\Psi^\alpha}^\dagger\circ \Psi^\alpha.
\]
The $r$th value of the axial current $j$ is given by 
\[
j_r=-e\,  \Psi^\dagger_r{^\alpha} \sigma_{\alpha\beta} \Psi_r^\beta,
\]
and the equation for the charge density is
\begin{equation}\label{kleingoru}
(\partial^2_0 \rho)_r-(D^2\rho)_r=-\frac{1}{2i}[(Dj)_r,\rho^t D^{-2}\rho].
\end{equation}
As in Ref. \cite{Cam07}, it is necessary to compute the Schwinger commutator $[j_j,\rho_{j'}]$ to obtain the asymptotic form of the mass term. To obtain  this commutator, we express it in terms of the $F$-transforms of $j_j$ and $\rho_{j'}$. We begin by  expanding $\Psi^{\alpha}_q$ and $\rho_q=e\Psi^\dagger_q \Psi_q$ in its $F$-components
\[
\Psi^{\alpha}_q=\sum_k[ a_ku^{\alpha}_k F_{qk} + b^\dagger_k{v^\dagger}^{\alpha}_k F^\dagger_{kq}],
\]
\[
\rho_q=e\sum_{k,k'}[ a^\dagger_ka_{k'}F^\dagger_{kq} F_{q{k'}} + b_kb^\dagger_{k'} F_{qk}F^\dagger_{{k'}q}].
\]
Thus, the $F$-transform $\tilde{\rho}_k=\sum_qF^\dagger_{kq}\rho_q$ becomes
\begin{equation}\label{roprdfs}
\tilde{\rho}_k=e\sum_{q,k',k''}[ a^\dagger_{k''}a_{k'}F^\dagger_{k''q}F_{q{k'}}F^\dagger_{kq} +
 b_{k''}b^\dagger_{k'}F_{q{k''}}F^\dagger_{{k'}q}F^\dagger_{{k}q}].
\end{equation}
Now, we have to take into account the explicit form of the matrix $F$. By using (\ref{vnkas}) and a trigonometric identity is possible to reduce the triple products appearing in (\ref{roprdfs}). For example we obtain that
\begin{equation}\label{tresefs}
F^\dagger_{k'' q}F_{qk'}F^\dagger_{k q}=\frac{1}{2N}\left(F^\dagger_{k''-k'+k,q}+F^\dagger_{k''+k'-k,q}-F^\dagger_{k''+k'+k,q}-F^\dagger_{k''-k'-k,q}
\right)
\end{equation}
and using (\ref{fasfg}), and the fact that $\Delta\kappa\simeq 1/N$,
\begin{eqnarray}\label{tresfs}
\sum_q F^\dagger_{k'' q}F_{qk'}F^\dagger_{k q}&=&\frac{\Delta\kappa}{2}(\delta_{\kappa'',\kappa'-\kappa}+\delta_{\kappa'',-\kappa'+\kappa}-\delta_{\kappa'',-\kappa'-\kappa}-\delta_{\kappa'',\kappa'+\kappa}\\
&-&\delta_{\kappa'',\kappa'-\kappa-1}-\delta_{\kappa'',-\kappa'+\kappa-1}+\delta_{\kappa'',-\kappa'-\kappa-1}+\delta_{\kappa'',\kappa'+\kappa-1}).\nonumber
\end{eqnarray}
Therefore, the $F$-transform $\tilde{\rho}_k$ of the charge density becomes 
\begin{eqnarray}\label{chdenk}
\tilde{\rho}_k&=&\frac{ae}{2\pi}\sum_{k'}(a^\dagger_{\kappa'-\kappa}a_{k'}+a^\dagger_{-\kappa'+\kappa}a_{k'}-a^\dagger_{-\kappa'-\kappa}a_{k'}-a^\dagger_{\kappa'+\kappa}a_{k'}\nonumber\\
&-&a^\dagger_{\kappa'-\kappa-1}a_{k'}-a^\dagger_{-\kappa'+\kappa-1}a_{k'}+a^\dagger_{-\kappa'-\kappa-1}a_{k'}+a^\dagger_{\kappa'+\kappa-1}a_{k'}\\
&+&b_{\kappa'-\kappa}b^\dagger_{k'}+b_{-\kappa'+\kappa}b^\dagger_{k'}-b_{-\kappa'-\kappa}b^\dagger_{k'}-b_{\kappa'+\kappa}b^\dagger_{k'}\nonumber\\
&-&b_{\kappa'-\kappa-1}b^\dagger_{k'}-b_{-\kappa'+\kappa-1}b^\dagger_{k'}+b_{-\kappa'-\kappa-1}b^\dagger_{k'}+b_{\kappa'+\kappa-1}b^\dagger_{k'}), \nonumber
\end{eqnarray}
where we have integrated the delta functions of argument $\kappa \pi/a$. The  $F$-transform $\tilde{j}_k$ of the axial current can be obtained similarly. In order to calculate $[\tilde{j}_k,\tilde{\rho}_{k'}]$, it is necessary to write this commutator in terms of normal-ordered operators with respect to the vacuum state filled with particles of any negative energy and antiparticles of any positive energy, which is defined by 
\[
\vert \text{vac}\rangle=\prod_{\kappa<0}a^\dagger_\kappa\prod_{\kappa>0}b^\dagger_\kappa\vert 0 \rangle.
\] 
The use of the anticommutation relations 
\begin{equation}\label{anticomopcd}
\{a^\dagger_\kappa,a_{\kappa'}\}=\delta_{\kappa\kappa'}, \quad \{b^\dagger_\kappa,b_{\kappa'}\}=\delta_{\kappa\kappa'},
\end{equation}
and a very lengthy calculation shows that the normal-ordered commutator $[\tilde{j}_k,\tilde{\rho}_{k'}]$ annihilates the vacuum state. In order to explain this, let us take only the first and third lines of (\ref{chdenk}) (corresponding to the first line of delta functions of (\ref{tresfs}) and to the delta centered at the origin of (\ref{fasfg})) and the analogous counterpart of $\tilde{j}_k$. These terms yield a contribution to the normal-ordered form of $[\tilde{j}_k,\tilde{\rho}_{k'}]$ acting on $\vert \text{vac}\rangle$ given by 
\begin{eqnarray}\label{sumrosin}
&&\frac{e^2a}{2\pi}\sum_m(a^\dagger_{\kappa-m} a_{-\kappa'-m}-a^\dagger_{-\kappa-m} a_{-\kappa'-m}+a^\dagger_{-\kappa+m} a_{-\kappa'-m}-a^\dagger_{\kappa+m} a_{-\kappa'-m}+a^\dagger_{-\kappa-m} a_{\kappa'-m}\nonumber\\
&-&a^\dagger_{\kappa-m} a_{\kappa'-m}-a^\dagger_{-\kappa+m} a_{\kappa'-m}+a^\dagger_{\kappa+m} a_{\kappa'-m}-a^\dagger_{-\kappa-m} a_{-\kappa'+m}+a^\dagger_{\kappa-m} a_{-\kappa'+m}\nonumber\\
&+&a^\dagger_{-\kappa+m} a_{-\kappa'+m}-a^\dagger_{\kappa+m} a_{-\kappa'+m}+a^\dagger_{-\kappa-m} a_{\kappa'+m}-a^\dagger_{\kappa-m} a_{\kappa'+m}-a^\dagger_{-\kappa+m} a_{\kappa'+m}\nonumber\\
&+&a^\dagger_{\kappa+m} a_{\kappa'+m}-b^\dagger_{-\kappa'-m} b_{-\kappa-m}+b^\dagger_{\kappa'-m} b_{-\kappa-m}-b^\dagger_{-\kappa'+m} b_{-\kappa-m}+b^\dagger_{\kappa'+m} b_{-\kappa-m}\\
&+&b^\dagger_{-\kappa'-m} b_{\kappa-m}-b^\dagger_{\kappa'-m} b_{\kappa-m}+b^\dagger_{-\kappa'+m} b_{\kappa-m}-b^\dagger_{\kappa'+m} b_{\kappa-m}+b^\dagger_{-\kappa'-m} b_{-\kappa+m}\nonumber\\
&-&b^\dagger_{\kappa'-m} b_{-\kappa+m}+b^\dagger_{-\kappa'+m} b_{-\kappa+m}-b^\dagger_{\kappa'+m} b_{-\kappa+m}-b^\dagger_{-\kappa'-m} b_{\kappa+m}+b^\dagger_{\kappa'-m} b_{\kappa+m}\nonumber\\
&-&b^\dagger_{-\kappa'+m} b_{\kappa+m}+b^\dagger_{\kappa'+m} b_{\kappa+m})\vert \text{vac}\rangle,\nonumber
\end{eqnarray}
where we have used the fact that $(ae/2\pi)^2Na/\pi=e^2a/2\pi$. Since $m$ takes negative and positive values, this expression vanish identically. The same situation occurs when the full expression of $\tilde{\rho}_k$ and $\tilde{j}_k$ are taken into account in the complete calculation of the commutator yielding that
\[
[\tilde{j}_k,\tilde{\rho}_{k'}]\vert \text{vac}\rangle=0.
\]
Thus, the inverse $F$-transform is also zero
\[
[j_k,\rho_{k'}]=0,
\]
and therefore, the right-hand side of (\ref{kleingoru}) vanishes
\begin{equation}\label{massterm}
m^2\rho_r=\frac{1}{2i}[(Dj)_r,\rho^t D^{-2}\rho]=0.
\end{equation}
This means that the bosonic mass associated to the Schwinger model is zero in this lattice formulation. This result is a consequence of the index  symmetry exhibited by (\ref{tresfs}), which comes from the fact that the diagonalizing matrix $F$ for the derivative (\ref{diffinsim}) satisfies relations like
(\ref{tresefs}), which, finally, are due to the specific form of $F$ and $D$. Note that this null result is independent of the number $N$ of lattice sites and the same result is obtained in the continuum limit. We give an explanation of this in the next section.
\section{Discussion}\label{secdis}
Firstly, let us consider a $N\times N$ matrix ${\cal D}$ very similar to (\ref{diffinsim}), which differs from $D$ only by the elements $D_{12}$ and $D_{21}$
\begin{equation}
{\cal D}=\frac{1}{2a} \begin{pmatrix} 0&\sqrt{2}&0&\cdots\\ \sqrt{2}&0&1&\cdots\\ 0&1&0\\ \vdots&\vdots&&\ddots\\ \end{pmatrix}.
\end{equation}
This matrix can also be considered as a derivative for a problem with homogeneous boundary conditions if $N$ is large. It is diagonalized by the unitary matrix which solves the eigenproblem ${\mathcal D} {\mathcal V}_k=i{\textsf p}_k {\mathcal V}_k$ \cite{Cam08}. The elements of ${\mathcal V}$ are given by
\begin{equation}\label{unkx}
({\mathcal V}_k)_q={\mathcal F}_{qk}=i^{q-1}\sqrt{\frac{2(1-{\textsf p}_k^2)}{N(1+\delta_{k0})}}\frac{T_{q-1}({\textsf p}_k)}{\vert T_{N-1}({\textsf p}_k)\vert},\quad q=1,2,\ldots,N,
\end{equation}
where ${\textsf p}_k=\eta_k/2a$, and $\eta_k$ is the $k$th zero of $N$th Tchebichef polynomial of the first kind $T_N(\eta)$. The asymptotic form of (\ref{unkx}) is
\begin{equation}\label{fcalqk}
{\mathcal F}_{qk}=(-i)^{q-1}\sqrt{\frac{2}{N(1+\delta_{k0})}}\cos\frac{(q-1)(k-1/2)\pi}{N},
\end{equation}
$q,k=1,2,\cdots, N$. The argument of the cosine function in (\ref{fcalqk}) takes the form $qk \pi/(N+1)\simeq qk \pi/N$ [the one of (\ref{vnkas})], if the zeros $\xi_k$ of $U_N(\xi)$ are used instead of $\eta_k$ to evaluate (\ref{unkx}).\\
Let us consider now an infinite lattice, i.e., a lattice with an infinite number of nodes and boundaries at infinity. Let $\delta$ be the central difference derivative [generalization of (\ref{diffinsim})].  Since there is not a first or last point in this lattice, we should write $\delta$ without reference to the first (or last) indexes
\[
\delta=\begin{pmatrix} \ddots\\ &0&1&0& \\ &-1&0&1& \\ &0&-1&0& \\ &&&&\ddots \end{pmatrix}.
\]
Any finite principal submatrix of $\delta$ coincides with the principal submatrices formed by deleting the first row and column of $D$ and ${\mathcal D}$ for some $N$, which become identical. Therefore, the asymptotic form of $F$ or the asymptotic form of ${\mathcal F}$, or both, can approximately diagonalize the matrix $\delta$, and a common choice is a linear combination of the sine and cosine functions [cf. Eqs. (\ref{vnkas}) and (\ref{fcalqk})] to form the usual discrete Fourier transform, eliminating the symmetry which gives a zero mass term (\ref{massterm}) for the Schwinger model in this lattice formulation and recovering the correct value for this term \cite{Cam07}. Note that any finite matrix with a nonzero border can be confused with $\delta$ in the asymptotic limit. This is the case of 
\begin{equation}\label{deldofer}
\Delta=\frac{1}{2a}\begin{pmatrix}
0&1&0&\cdots&0&\mp 1\\
                      -1&0&1&\cdots&0&0\\
                      0&-1&0&\cdots&0&0\\
                      \vdots&\vdots&\vdots&\ddots&\vdots&\vdots\\
                      0&0&0&\cdots&0&1\\
                      \pm 1&0&0&\cdots&-1&0\\
\end{pmatrix},
\end{equation}
where $\pm$ is discriminated according to the parity of $N$. This matrix is diagonalized by the discrete Fourier transform $\exp(-i 2\pi j k/N)/\sqrt{N}$ and their eigenvalues can be ordered as 
\[
\frac{i}{a}\sin \frac{2\pi j}{N},
\]
where the index $j$ runs over integers or half-integers \cite{Cam01b}. This is the matrix which yields the fermion doubling since it bends the Brillouin zone onto itself.\\
It is worth to notice that Eq. (\ref{tresefs}) plays an important role in the calculation of the mass term and to remind that the matrix $F$ is determined by the boundary conditions. For example, for the usual discrete Fourier transform, there is only one term in the right-hand side of (\ref{tresefs}). As another example, we can take the matrices $({\mathcal F},{\cal D})$ instead $(F,D)$ and calculate again the mass term. In this case, Eq. (\ref{tresefs}) becomes
\[
{\cal F}^\dagger_{k'' q}{\cal F}_{qk'}{\cal F}^\dagger_{k q}=\frac{1}{2N}\left({\cal F}^\dagger_{k''-k'+k,q}+{\cal F}^\dagger_{k''+k'-k,q}+{\cal F}^\dagger_{k''+k'+k,q}+{\cal F}^\dagger_{k''-k'-k,q}\right)
\]
and taking into account that the value $m=0$ should be excluded from the sums since the number of nodes $N$ is an even number, Eq. (\ref{sumrosin}) changes to an expression that does not vanish identically:
\begin{eqnarray}\label{sumrocos}
&&\frac{e^2a}{\pi}(\delta_{k'k}+\delta_{-k',k}){\sum_m} \Big[ (a^\dagger_m a_m - a^\dagger_{-\kappa-m} a_{-\kappa-m}) + 
 (a^\dagger_m a_m - a^\dagger_{\kappa-m} a_{\kappa-m})\nonumber\\
&+& (b^\dagger_m a_m - b^\dagger_{-\kappa-m} a_{-\kappa-m}) + 
(b^\dagger_m a_m - b^\dagger_{\kappa-m} a_{\kappa-m})\Big]\vert \text{vac}\rangle.\nonumber
\end{eqnarray}
The $a$-terms of the sum applied to $\vert \text{vac}\rangle$ produce
\[
\sum_m (a^\dagger_m a_m - a^\dagger_{-\kappa-m} a_{-\kappa-m})\vert \text{vac}\rangle=-k\vert \text{vac}\rangle,\quad 
\sum_m (a^\dagger_m a_m - a^\dagger_{\kappa-m} a_{\kappa-m})\vert \text{vac}\rangle=k\vert \text{vac}\rangle,
\]
giving a zero contribution to the mass term. The same result is obtained for the $b$-terms of the sum and the total contribution to the term mass is zero again.\\
Let us remark finally that for a {\it finite} lattice only one diagonalizing matrix for the derivative $D$ (or ${\cal D}$) exists, and therefore, there is no way to get rid of the null result which becomes a finite-size effect. 

\section{Acknowledgment}
The authors want to thank the referee for calling our attention to the fermion doubling problem and to discuss in more detail some results.

\end{document}